# A NEW PERTURBATION TECHNIQUE FOR EIGENENERGIES OF THE SCREENED COULOMB POTENTIAL

**I.V. Dobrovolska, R. S. Tutik**

Dniepropetrovsk National University, Department of Physics,
Naukova str., 13, 49050, Dniepropetrovsk, Ukraine
e-mail: tutik@ff.dsu.dp.ua

*Abstract* – **The explicit semiclassical treatment of the logarithmic perturbation theory for the bound-state problem of the radial Shrödinger equation with the screened Coulomb potential is developed. Based upon $\hbar$-expansions and new quantization conditions a novel procedure for deriving perturbation expansions is offered. Avoiding disadvantages of the standard approach, new handy recursion formulae with the same simple form both for ground and excited states have been obtained.**

## I. INTRODUCTION

The problem of accurately determining the properties of energy eigenstates of the screened Coulomb potential has been considerable interest in numerous areas of physics for many years. The Shrödinger equation with this potential is not exactly solvable and numerical and approximate methods have been applied to obtain the energy eigenvalues. Despite a variety of approximation methods one of the most popular approaches is still the logarithmic perturbation theory [1-4]. Within the framework of this theory, the conventional way to solve the bound-state problem consists in changing from the wave function to its logarithmic derivative. In the case of ground states, the consequent expansion in a small parameter leads to a hierarchy of simple equations that permits us to derive easily the corrections to the energy and wave functions for each order. However, when radially excited states are considered, the standard approach [4] becomes extremely cumbersome and, practically, inapplicable. At the same time the evaluation of perturbation terms of large orders is needed for applying modern summation procedures because the obtained series are typically divergent.

Recently, a new technique based on specific quantization conditions has been proposed to get the perturbation series via the semiclassical $\hbar$-expansions within the framework of the one dimensional Shrödinger equation [5]. Avoiding disadvantages of the standard approach this straightforward semiclassical procedure results in new handy recursion formulae with the same simple form both for ground and excited states. Moreover, these formulae can be easily applied to any renormalization scheme of improving the convergence of expansions [6].

The object of this paper is to extend the above mentioned formalism to the bound-state problem for the radial Shrödinger equation with the screened Coulomb potential.

## II. THE METHOD

We study the bound-state problem with a discrete energy spectrum for a non-relativistic particle moving in the central field of a static screened Coulomb potential.

To construct the method we start with the radial Shrödinger equation

$$-\frac{\hbar^2}{2m}U''(r)+V(r)U(r)+\frac{\hbar^2 l(l+1)}{2m r^2}U(r)=EU(r), \qquad (1)$$

where the potential $V(r)$ has a Coulomb-like behavior at the origin.

In common practice, the screened Coulomb potential has a form $V(r)=r^{-1}f(\lambda,r)$, where $\lambda$ is a small parameter. With the standard approach, the screening function, $f(\lambda,r)$, is expanded in the Taylor series involving in the consideration the powers of the screening parameter $\lambda$ in combination with derivatives of the screening function. However, on performing the scale transformation $r \to \hbar^2 r$ it is seen that the powers of the screening parameter appear in common with powers of Planck's constant squared. Hence it appears that the perturbation series must be in reality not only $\lambda$-expansions but also the semiclassical $\hbar^2$-expansions, too.

Thus, as stated earlier, we intend to restore results of the logarithmic perturbation theory by means of the explicit expansions in powers of $\hbar$. Besides, in what follows we do not single out explicitly the screening parameter, but incorporate it into constants $V_i$ of the expansion of the potential $V(r) = r^{-1} \sum_{i=0}^{\infty} V_i r^i$.

As is customary in the logarithmic perturbation theory, we go over from the wave function, $U(r)$, to its logarithmic derivative, $C(r) = \hbar U'(r)/U(r)$, and arrive at the Riccati equation

$$\hbar C'(r) + C^2(r) = \frac{\hbar^2 l(l+1)}{r^2} + 2mV(r) - 2mE. \tag{2}$$

Notice, that in the complex plane, a number of zeros $N$ of the wave function $U(r)$ inside the closed contour is defined by its logarithmic derivative through the quantization condition

$$\frac{1}{2\pi i} \oint_\gamma C(r) dr = \hbar N, \tag{3}$$

that is known as the principle of argument in the analysis of complex variables.

This condition must be supplemented with a rule of achieving a classical limit for radial ($n$) and orbital ($l$) quantum numbers which are the specific quantum notions. The semiclassical treatment of the logarithmic perturbation theory proved to involve the rule

$$\hbar \to 0, \; n = const, \; l = const, \; \hbar n \to 0, \; \hbar l \to 0. \tag{4}$$

It follows from this rule that in the classical limit, as $\hbar \to 0$, the centrifugal term in (1) disappears and a particle falls into center. Then the classical turning points, i.e. zeros of the classical momentum, draw to the origin and the nodes of the wave function are joined together at $r = 0$. Hence, the contour of integration in the quantization condition must enclose only the origin. The multiplicity of zero formed at this point includes the number of nodes and the value $l+1$ caused by the requirement of regular behavior of the wave function. Therefore the total number of zeros in the quantization condition (3) becomes equal to *N=n+l+1*.

We assume that the energy eigenvalues and the logarithmic derivative of the wave function may be written as an asymptotic power series in the Planck constant squared. After deriving the leader order of them from the Riccati equation (2) we have

$$E = \hbar^{-2} \sum_{k=0}^{\infty} E_k \hbar^{2k}, \; C(r) = \hbar^{-1} \sum_{k=0}^{\infty} C_k(r) \hbar^{2k}. \tag{5}$$

Then the quantization condition (3) is finally rewritten as

$$\frac{1}{2\pi i} \oint_\gamma C_i(r) dr = (n+l+1)\delta_{i,1}. \tag{6}$$

A further application of the theorem of residues to the explicit form of functions $C_k(r)$ easily solves the problem of the description of both ground and excited states.

III. RECURSION RELATIONS

We proceed now to deriving the recursion relations for calculation of the *n*th eigenfunction and corresponding energy eigenvalue.

Upon inserting (5) into (2) and equating coefficients of equal powers of $\hbar$ to zero, we obtain

$$C_0^2 = -2mE_0, \quad C_0(r)C_1(r) = m[V(r) - E_1],$$

$$C_1' + 2C_0(r)C_2(r) + C_1^2(r) = \frac{l(l+1)}{r^2} - 2mE_2, \quad (7)$$

$$C_{k-1}'(r) + \sum_{j=0}^{k} C_j(r) C_{k-j}(r) = -2mE_k, \quad k > 2.$$

From this system we can see that $C_1(r)$ has a simple pole at the origin, owing to the Coulomb behavior of the potential at this point, while the function $C_k(r)$ has a pole of the order $k$. Hence $C_k(r)$ can be represented by the Laurent series

$$C_k(r) = r^{-k} \sum_{i=0}^{\infty} C_i^k r^i, \quad k \geq 1. \quad (8)$$

The substitution of this expansion into the system and collecting coefficients of the like powers of $r$ leads to the following recursion relations in terms of the Laurent coefficients, $C_i^k$:

$$C_0^0 = -\sqrt{-2mE_0}, \quad C_i^0 = -\frac{1}{2C_0^0}[V_i - E_1\delta_{i,1}], \quad E_0 = -\frac{mV_0^2}{2(n+l+1)^2}, \quad E_1 = V_1,$$

$$C_i^k = -\frac{1}{2C_0^0}\Big[(i-k+1)C_i^{k-1} + \sum_{j=1}^{k-1}\sum_{p=0}^{i} C_p^j C_{i-p}^{k-j} + 2mE_k\,\delta_{i,k} - l(l+1)\delta_{i,0}\,\delta_{k,2}\Big], \quad k > 1, \quad (9)$$

$$C_k^{k-1} = (n+l+1)\delta_{k,1}, \quad E_k = -\frac{1}{2m}\Big[C_k^{k-1} + \sum_{j=1}^{k-1}\sum_{p=0}^{k} C_p^j C_{k-p}^{k-j} + 2C_0^0 C_k^k\Big], \quad k > 1.$$

Thus, the problem of the same description of the ground and excited states can be considered solved.

IV. Conclusion

In this paper, a new useful semiclassical technique for deriving results of the logarithmic perturbation theory for the bound-state problem within the framework of the radial Shrödinger equation with the screened Coulomb potential has been developed. Based upon the $\hbar$-expansions and new quantization conditions, novel handy recursion relations have been obtained. Avoiding the disadvantages of the standard approach these formulae have the same simple form both for ground and exited states and provide, in principle, the calculation of the perturbation corrections up to an arbitrary order in the analytic or numerical form.